\documentclass[twocolumn,preprintnumbers,amsmath,amssymb,superscriptaddress]{revtex4}

\usepackage{graphicx}
\usepackage{graphicx}
\begin{document}
\title{Disappearance of zinc impurity resonance in large gap region on Bi$_{\mathrm{2}}$Sr$_{\mathrm{2}}$CaCu$_{\mathrm{2}}$O$_{\mathrm{8+}\delta}$ probed by scanning tunneling spectroscopy}

\author{Tadashi Machida}
\affiliation{Superconducting Materials Center, National Institute for Materials Science, 1-2-1 Sengen, Tsukuba, Ibaraki 305-0047}

\author{Takuya Kato}
\affiliation{Department of Physics, Tokyo University of Science, 1-3 Kagurazaka, Shinjuku-ku, Tokyo 162-8601, Japan}

\author{Hiroshi Nakamura}
\affiliation{Department of Physics, Tokyo University of Science, 1-3 Kagurazaka, Shinjuku-ku, Tokyo 162-8601, Japan}

\author{Masaki Fujimoto}
\affiliation{Department of Physics, Tokyo University of Science, 1-3 Kagurazaka, Shinjuku-ku, Tokyo 162-8601, Japan}

\author{Takashi Mochiku}
\affiliation{Superconducting Materials Center, National Institute for Materials Science, 1-2-1 Sengen, Tsukuba, Ibaraki 305-0047}

\author{Shuuichi Ooi}
\affiliation{Superconducting Materials Center, National Institute for Materials Science, 1-2-1 Sengen, Tsukuba, Ibaraki 305-0047}

\author{Ajay D. Thakur}
\affiliation{Superconducting Materials Center, National Institute for Materials Science, 1-2-1 Sengen, Tsukuba, Ibaraki 305-0047}

\author{Hideaki Sakata}
\affiliation{Department of Physics, Tokyo University of Science, 1-3 Kagurazaka, Shinjuku-ku, Tokyo 162-8601, Japan}

\author{Kazuto Hirata}
\affiliation{Superconducting Materials Center, National Institute for Materials Science, 1-2-1 Sengen, Tsukuba, Ibaraki 305-0047}
\date{\today}
\begin{abstract}

Using Scanning tunneling spectroscopy (STS), we report the correlation between spatial gap inhomogeneity and the zinc (Zn) impurity resonance in single crystals of Bi$_{\mathrm{2}}$Sr$_{\mathrm{2}}$Ca(Cu$_{\mathrm{1-}x}$Zn$_{x}$)$_{\mathrm{2}}$O$_{\mathrm{8+}\delta}$ with different carrier (hole) concentrations ($p$) at a fixed Zn concentration ($x$ $\sim$ 0.5 \% per Cu atom). In all the samples, the impurity resonance lies only in the region where the gap value is less than $\sim$ 60 meV. Also the number of Zn resonance sites drastically decreases with decreasing $p$, in spite of the fixed $x$. These experimental results lead us to a conclusion that the Zn impurity resonance does not appear in the large gap region although the Zn impurity evidently resides in this region.

\end{abstract}
\maketitle
The effect of a single impurity in high temperature superconductors (HTSCs) with a $d$-wave pairing symmetry is one of the most intriguing issues in the physics of correlated electron systems and it has attracted a lot of theoretical\cite{Baratsky_1,Baratsky_2,Kruis,Tang_1,Tang_2,Andersen_1,Polkovnikov,Kircan,Vojta,Andersen_Phase_1} as well as experimental\cite{Hudson_1,Pan_1,Hudson_2,Chatterjee,Kambara,Lang,Hoffman,Machida_2} attention in recent past. The impurity atoms significantly modify the bulk superconducting properties and it has been anticipated that the impurity can be used to understand the nature of local superconductivity around the impurity. Experimentally, the scanning tunneling spectroscopy (STS) measurements based on scanning tunneling microscope (STM) have given us the detailed information on the local density-of-states (LDOS) in the vicinity of a single non-magnetic\cite{Pan_1} (Zn) and magnetic\cite{Hudson_2} (Ni) impurities in Bi$_{\mathrm{2}}$Sr$_{\mathrm{2}}$CaCu$_{\mathrm{2}}$O$_{\mathrm{8+}\delta}$ (Bi2212) via the local measurements of the tunneling conductance $dI/dV = g(\mathrm{\textbf{r}}, V)$. Particularly, in the case of Zn, each impurity creates a sharp conductance peak lying close to the Fermi energy ($E_{\mathrm{F}}$) near the Zn site with the maximum intensity at Zn site and the second largest intensity on the four next-nearest neighbor sites\cite{Pan_1}. There are several theoretical scenarios that ascribe the conductance peak near $E_{\mathrm{F}}$ to : (i) the electronic scattering by impurities with non-magnetic scattering potentials\cite{Baratsky_1,Baratsky_2,Kruis,Tang_1,Tang_2}, (ii) the local sign change in the superconducting order parameter near the impurity site\cite{Andersen_1,Andersen_Phase_1}, and (iii) to the Kondo screening of a local spin induced by non-magnetic impurity\cite{Polkovnikov,Kircan,Vojta}. However, no consensus has yet been reached about the proper explanation of the conductance peak near $E_{\mathrm{F}}$ induced by the Zn impurity.

\begin{figure}[tb]
\begin{center}
\includegraphics[width=5.7cm]{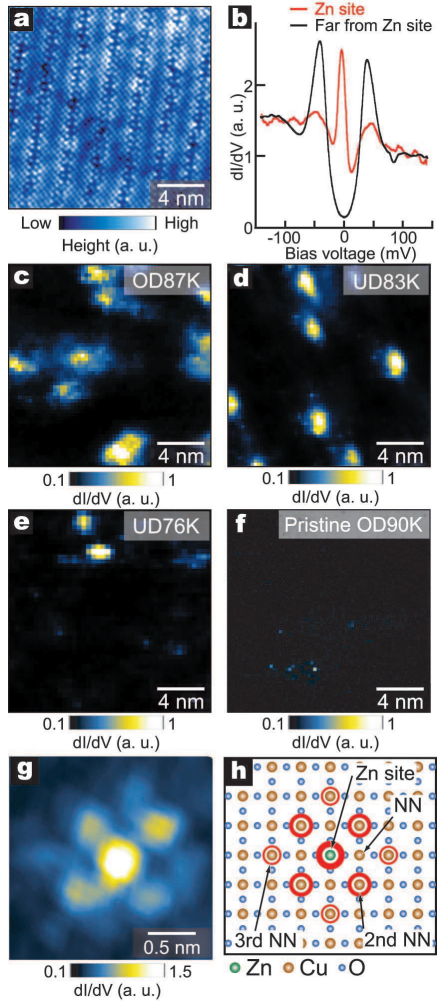}
\end{center}
\caption{(Color online) \textbf{a}, Typical 17 $\times$ 17 nm$^{\mathrm{2}}$ STM image on a BiO surface of Bi2212 taken at the sample bias voltage ($V_{\mathrm{B}}$) of 300 mV and the setup current ($I_{\mathrm{set}}$) of 100 pA. \textbf{b}, Typical tunneling spectra taken at a Zn site (red) and at a site far from Zn site (black) which are scaled by the conductance at the $V_{\mathrm{B}}$ = +100 mV. \textbf{c} - \textbf{f} Typical normalized conductance maps at -1.5 mV [$g(\mathrm{\textbf{r}}, \mathrm{-1.5 mV})$/$g(\mathrm{\textbf{r}}, \mathrm{+100 mV})$] taken on 17 $\times$ 17 nm$^{\mathrm{2}}$ regions in four samples which are OD87K (\textbf{c}), UD83K (\textbf{d}), UD76K (\textbf{e}), and Pristine (\textbf{f}), respectively. In these maps, the contrast has been fixed for the comparison among them. \textbf{g} A high spatial resolution $g(\mathrm{\textbf{r}}, \mathrm{-1.5 mV})$ map around a Zn site. \textbf{h} A schematic illustration of the CuO$_{2}$ plane. The sites showing the local maximum in the conductance map in \textbf{g} marked by red(bold) circles in \textbf{h}. These figures indicate that the local maximum appear at the Zn, second nearest neighbor (2nd NN), and third nearest neighbor (3rd NN) sites.}
\label{Fig_Gmap}
\end{figure}

In addition, the STS measurements have revealed the existence of an inhomogeneous energy gap ($\Delta$) distribution on nanometer length scales, which is commonplace in Bi-based superconductors \cite{Lang,Hoffman,Pan_2,McElroy_1,Howald,Gomes,Fang,Machida_1,Kato}. Lang \textit{et al.}\cite{Lang} and Hoffman \textit{et al.}\cite{Hoffman} have studied the spatial correlation between the impurity resonance and the gap inhomogeneity.
In their experiments, the impurity resonance has been observed in the small gap region with $\Delta \le$ 50 meV. This in turn suggests the following interpretations : (i) the impurity resonances are observable at all the impurity sites, and the local $\Delta$ is suppressed near the impurity sites, or (ii) the impurity resonance is not observable in the large gap region with $\Delta \ge$ 50 meV, although the impurities reside in that region. There has not been a dedicated experimental investigation indicating which of these two interpretations is more suitable. Nevertheless, several theoretical studies regarding the absence of the impurity resonance in the large gap region have been performed by adopting the latter interpretation \cite{Andersen_1,Kircan}. It is therefore of fundamental interest to experimentally establish the correct physical picture for the observation of Zn resonances. This could be achieved by the investigation of the spatial correlations between the impurity resonance and the gap inhomogeneity using samples with the same impurity concentration but with different carrier (hole) concentrations ($p$). This in turn is motivated by the observations that the large gap region becomes dominant with decreasing $p$\cite{Lang,Hoffman,McElroy_1,Gomes}. For instance, if the former interpretation is valid, the number of impurity resonances does not change with changing $p$, whereas, if the latter is valid, the number of these resonances decreases with decreasing $p$ (with corresponding increase in large gap regions). In this {\it rapid communication}, we report our investigations on the spatial correlation between the gap inhomogeneity and the Zn impurity resonance sites in Bi$_{\mathrm{2}}$Sr$_{\mathrm{2}}$Ca(Cu$_{\mathrm{1-}x}$Zn$_{x}$)$_{\mathrm{2}}$O$_{\mathrm{8+}\delta}$ (Zn-doped Bi2212) with different $p$ and the same Zn concentration ($x$) by using the STS technique. The results strongly suggests the validity of the latter scenario, where, the impurity resonance is not observable in the large gap regions although the impurities reside there.

The samples used in this study are the floating-zone grown single crystals of Zn-doped Bi2212 with different $p$ controlled by oxygen depletion. The value of $x$ was determined to be $\sim$ 0.5\% by inductively-coupled plasma optical emission spectrometry after the crystallization (nominal concentration $\sim$ 2\%). We prepared three samples whose carrier concentrations were : (i) slightly overdoped (as-grown), (ii) slightly underdoped, and (iii) underdoped with superconducting transition temperatures ($T_{\mathrm{c}}$) of 87 K, 83 K, and 76 K, respectively (as determined by SQUID magnetization measurements).  We labeled these samples as OD87K, UD83K, and UD76K, respectively. A home-built low temperature STM was used for the STS measurements. All STS measurements were performed in a helium gas environment at 4.2 K on an atomically clean and flat surface as shown in Fig. \ref{Fig_Gmap}\textbf{a} prepared by cleaving the samples \textit{in situ} at 4.2 K in pure helium gas.

To visualize the Zn resonance sites, we mapped out the conductance at -1.5 mV in each sample as shown in Figs. \ref{Fig_Gmap}\textbf{c}-\textbf{e} (since the resonance peak in LDOS appears near the energy of -1.5 meV at Zn site as shown in Figs. \ref{Fig_Gmap}\textbf{b} and already reported elsewhere\cite{Pan_1,Hoffman}).
In these maps, there are several bright sites in an overall dark background, whereas, there is no clear bright site in the map of the pristine sample as shown in Figs. \ref{Fig_Gmap}\textbf{f}.
The four-fold symmetric structure near the resonance sites is not visible in Figs. \ref{Fig_Gmap}\textbf{c}-\textbf{e}.
This is merely due to the low spatial resolution, because the four-fold structure is clearly visible in Fig. \ref{Fig_Gmap}\textbf{g} which is a high spatial resolution image near a representative Zn impurity.
Therefore, it is plausible to consider that the observed bright sites in Figs. \ref{Fig_Gmap}\textbf{c}-\textbf{e} correspond to the sites showing the Zn impurity resonance.
The number of the bright sites corresponding to the Zn impurity resonance site decreases with diminishing $p$ (particularly this feature is remarkable for UD76K), although $x$ is fixed in all samples. This indicates that there are some regions in which the Zn resonance vanishes and that such regions gradually expand with decreasing $p$.

\begin{figure}[tb]
\begin{center}
\includegraphics[width=6cm]{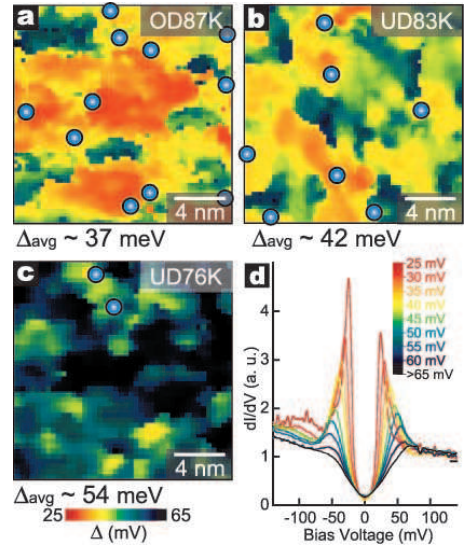}
\end{center}
\caption{(Color) \textbf{a} - \textbf{c} Gap maps on the same regions as in Fig.1\textbf{c} - \textbf{e}, respectively. The sites showing Zn resonance are superimposed on the gap maps as the light blue spheres. \textbf{d} The average spectrum associated with each $\Delta$, where the color of each gap-average spectrum corresponds to the color scale used in all gap maps. The spectra are normalized by the conductance value at +100 mV.}
\label{Fig_Gap}
\end{figure}

Figures \ref{Fig_Gap}\textbf{a}-\textbf{c} show the gap maps ($\Delta$ map) on the same field of view as shown in Figs. \ref{Fig_Gmap}\textbf{c}-\textbf{e}. The gap values are spatially distributed from 25 mV to 55 mV for OD87K, from 25 mV to 75 mV for UD83K, and from 35 to 85 mV for UD76K, respectively. Figure \ref{Fig_Gap}\textbf{d} shows the average spectrum associated with each local $\Delta$ in all maps and reveals a substantial variation of the spectrum shape changing local $\Delta$ as follows : (i) the peak heights of the gap edges considerably fall off with increasing local $\Delta$ and (ii) the peak structure at negative gap edge becomes ill-defined in the spectra with $\Delta$ $\ge$ 60 mV.
The region showing the spectrum with the ill-defined peak (dark region in $\Delta$ maps) becomes dominant with decreasing $p$.
Consequently, the spatially averaged $\Delta$ value also increases with decreasing $p$. These features regarding the gap inhomogeneity are in good agreement with those reported in previous studies \cite{Lang,Hoffman,Pan_2,McElroy_1,Gomes,Machida_1,Kato}. We superimposed the light blue spheres on these gap maps as the Zn resonance sites to clarify the spatial correlation between the local $\Delta$ and the Zn resonance. Apparently, the Zn resonance occurs only in the regions with the small gap value less than about 60 meV.

\begin{figure}[tb]
\begin{center}
\includegraphics[width=6cm]{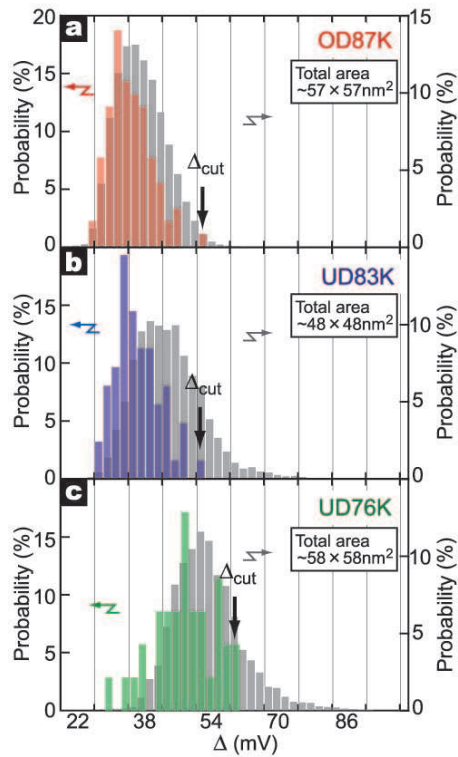}
\end{center}
\caption{(Color) Statistical analysis of local $\Delta$ on OD87K \textbf{a}, UD83K \textbf{b}, and UD76K \textbf{c}, respectively.. In these analyses, we collected both the data in the regions shown in Fig. 1 and those in other regions. The total areas used in the analyses are 57 $\times$ 57 nm$^{\mathrm{2}}$ (OD87K \textbf{a}), 48 $\times$ 48 nm$^{\mathrm{2}}$ (UD83K \textbf{b}), and 59 $\times$ 59 nm$^{\mathrm{2}}$ (UD76K \textbf{c}), respectively. Gray bars in all figures show the histograms of the local $\Delta$ in overall regions including the site showing the Zn resonance. On the other hand, the colored bars show the histograms of the local $\Delta$ at only the sites showing the Zn resonance. We determined the number of the Zn resonance sites per Cu atom to be 0.42 (OD87K \textbf{a}), 0.37  (UD83K \textbf{b}), and 0.15 \% (UD76K \textbf{c}), respectively, by counting the resonance sites in all regions.}
\label{Fig_Hist}
\end{figure}

To improve the reliability of the spatial correlation, we performed the same experiments on several regions of the samples. Also the total scanning area was 57 $\times$ 57 nm$^{2}$ for OD87K,  48 $\times$ 48 nm$^{2}$ for UD83K, and  59 $\times$ 59 nm$^{2}$ for UD76K, respectively.  A statistical analysis of the collected data was performed. In this analysis, we picked up a large number of the Zn resonance sites whose average concentrations (the total number of the sites) were 0.42 \% per Cu (95 counts) for OD87K, 0.37 \% (60 counts) for UD83K, and 0.15 \% (36 counts) for UD76K, respectively. Fig. \ref{Fig_Hist} shows the histograms of the gap values taken in all regions (gray bar) and at the Zn resonance sites (colored bar) in each sample.
The maximum gap value in the colored histograms ($\Delta_{\mathrm{cut}}$) indicates that there is no Zn resonance site in the region with the $\Delta$ larger than $\Delta_{\mathrm{cut}}$. In overdoped sample, $\Delta_{\mathrm{cut}}$ is very close to the maximum gap value ($\Delta_{\mathrm{max}}$) in the gray histogram obtained from all the regions. For underdoped samples, the values of $\Delta_{\mathrm{cut}}$ are clearly smaller than $\Delta_{\mathrm{max}}$. This means that the Zn resonance can be observed almost in the entire area in overdoped sample, whereas it partially appears in the region with $\Delta$ $\le$ $\Delta_{\mathrm{cut}}$ in the underdoped samples. From the above systematic statistical analysis, it has been revealed that : (i) the Zn resonance appears only in the regions with $\Delta \le $ $\Delta_{\mathrm{cut}}$, and (ii) the number of the resonance sites decreases with the reduction of $p$. These two results clearly validate the viewpoint that the Zn resonance does not arise in the large gap region with $\Delta \ge $ $\Delta_{\mathrm{cut}}$ although the Zn atoms really resides in such large gap regions.

These experimental data raise a pertinent question : ``what is the important factor governing the formation of the Zn resonance peak?''
As a matter of course, our data imply the possibility that the formation of the resonance peak is governed by gap size itself.
Here, we discuss other possibilities.
The suggestion for the absence of the impurity resonance in large gap region was reported by Lang \textit{et al.} who insisted that the superconductivity is suppressed in the large gap region and that superconductivity is crucial for the formation of the resonance\cite{Lang}. Recently, the presence of the resonance above $T_{\mathrm{c}}$, however, has been experimentally observed by Chatterjee \textit{et al.}\cite{Chatterjee}. Their results mean that the superconductivity is not essential in the formation of the resonance.
Furthermore, it has been established that $p$ in large gap region is lower than that in small gap region from the local quasiparticle interference information\cite{Wise}. This enables us to interpret the spatial variation of $\Delta$ as that of $p$. Therefore, the absence of the Zn resonance in large gap region can be interpreted as that in low-$p$ region.
From above discussion, it is also highly likely that the carrier density governs the formation of the impurity resonance.
Thus, both the gap size and the carrier density can be the key parameter.
Unfortunately, it is difficult to determine which is more crucial, since these parameters are experimentally equivalent quantities.

In terms of theoretical studies on the absence of the Zn resonance in large gap regions, there are three scenarios as follows: (i)the simple screened Coulomb impurity scenario\cite{Andersen_1}, (ii) the Kondo impurity scenario\cite{Kircan}, and (iii) the phase impurity scenario\cite{Andersen_1,Andersen_Phase_1}.
All of the results in these works indicate that the enhancement of the local gap size suppresses the impurity resonance peak.
These theoretical studies have focused on the gap size as the key parameter regarding the formation of the impurity resonance.
However, there are few theoretical works focusing on the carrier density around the impurity. Since our results imply a possibility that the carrier density can be the key factor for the formation of the Zn resonance, it appears to be important that the carrier density is taken into account as one of the crucial keys in discussing the origin of the Zn resonance peak.

In summary, we have performed scanning tunneling spectroscopy (STS) measurements on the single crystals of Bi$_{\mathrm{2}}$Sr$_{\mathrm{2}}$Ca(Cu$_{\mathrm{1-}x}$Zn$_{x}$)$_{\mathrm{2}}$O$_{\mathrm{8+}\delta}$ with the different carrier concentrations $p$, keeping the Zn concentration $x$ $\sim$ 0.5 \% per Cu atom. The absence of the Zn resonance in the large gap region with the gap value more than 60 meV have been found in all samples.
In addition to this correlation, the significant reduction in the number of Zn resonance sites with decreasing $p$ can be observed in spite of the fixed $x$. These results provide a conclusion that the impurity resonance does not occur in the large gap region even though the Zn atom actually exists in this region. This conclusion gives us the possibility that the local gap size governs the formation of the Zn resonance. Furthermore, from comparison with previous STS studies, it is also plausible that not only the gap size but also the carrier density around the impurity (rather than superconductivity) governs the formation of Zn resonance. This will be a crucial key for figuring out a long-standing puzzle with respect to the origin of Zn impurity resonance.


\end{document}